\documentclass[12pt]{iopart}

\usepackage{epsfig}
\usepackage{color}
\usepackage{graphics}
\usepackage{amssymb}
\usepackage{iopams}

\begin{document}

\title[Facilitation of polymer looping in baths of active particles]{Facilitation
of polymer looping and giant polymer diffusivity in crowded solutions of active
particles}

\author{Jaeoh Shin $^{\dagger,\ddagger}$, Andrey G. Cherstvy$^\dagger$, W. K.
Kim$^\flat$, and Ralf Metzler$^{\dagger,\sharp,1}$}
\address{$^\dagger$Institute for Physics \& Astronomy, University of Potsdam,
14476 Potsdam-Golm, Germany\\
$^\ddagger$ Max Planck Institute for the Physics of Complex Systems,
01187 Dresden, Germany\\
$\flat$ Fachbereich Physik, Freie Universit{\"a}t Berlin, 14195 Berlin, Germany\\
$^\sharp$Department of Physics, Tampere University of Technology, 33101
Tampere, Finland}
\ead{$^1$rmetzler@uni-potsdam.de}

\date{\today}

\begin{abstract}
We study the dynamics of polymer chains in a bath of self-propelled particles
(SPP) by extensive Langevin dynamics simulations in a two dimensional system.
Specifically, we analyse the polymer looping properties versus the SPP
activity and investigate how the presence of the active particles alters the
chain conformational statistics. We find that SPPs tend to extend flexible
polymer chains while they rather compactify stiffer semiflexible polymers, in
agreement with previous results. Here we show that larger activities of SPPs
yield a higher effective temperature of the bath and thus facilitate looping
kinetics of a passive polymer chain. We explicitly compute the looping
probability and looping time in a wide range of the model parameters. We
also analyse the motion of a monomeric tracer particle and the polymer's
centre of mass in the presence of the active particles in terms of the time
averaged mean squared displacement, revealing a giant diffusivity enhancement
for the polymer chain via SPP pooling. Our results are applicable to rationalising
the dimensions and looping kinetics of biopolymers at constantly fluctuating
and often actively driven conditions inside biological cells or suspensions
of active colloidal particles or bacteria cells.
\end{abstract}

\section{Introduction}
\label{intro}

Active motion is a necessary prerequisite for living systems to maintain vital
processes, including materials transport inside cells and the foraging dynamics
of mobile organisms \cite{review12b,review13}. The length scales associated with
active motion processes span several orders of magnitude and range from the
nanoscopic motion of cellular molecular motors \cite{julicher-motors} essential
to move larger cargo in the crowded environment of cells \cite{igor}, over the
microscopic motion of bacteria cells and micro-swimmers \cite{elgeti-microswimmers,
review12}, to the macroscopic motion patterns of higher animals and
humans \cite{peshehody}. In particular, artificial Janus colloids are propelled
by diffusiophoretic or thermophoretic forces \cite{golestanian07,sana10,peter}.
Active motion enhances the speed and precision of signalling and cargo
transport in biological cells \cite{aljaz} and allows efficient search of sparse
targets for large organisms \cite{search}.

A somewhat different question is how passive particles are influenced by an active
environment. Tracking the motion of tracer particles immersed in baths of active
bacteria \cite{lipchaber00, poon13} and swimming eukaryotic cells \cite{goldstein09}
one typically observes an enhanced effective tracer mobility, and the active
environment may lead to exponential tails of the displacement distribution
\cite{goldstein09,marenduzzo14}. Passive particles may also become enslaved to
the motion of motor-cargo complexes due to cytoplasmic drag \cite{haim}. When
micron sized colloids are immersed in baths with smaller particles, short ranged
attractive depletion forces of entropic origin are observed \cite{ao58}. However,
the same colloids in a bath of self-propelled particles (SPPs) may experience long
ranged attractive or repulsive forces depending on the SPP characteristics
\cite{cacciuto14depletion}. By tuning the concentration of SPPs the forces between
two plates can be controlled \cite{ni15}.

Here we want to focus on the properties of polymer chains in an active liquid. It
is known that when a polymer chain is immersed into an SPP bath its extension
changes non-monotonically with the activity $F_a$ of active particles due to
competing effects of active forces and chain elasticity \cite{kaiser14,cacciuto14}.
We study here the extent to which the activity of SPPs alters the \emph{internal}
motion of a polymer chain, specifically, how its end loop formation kinetics becomes
affected. Polymer looping or reactions of the chain ends is a fundamental process
governing numerous biological functions \cite{looping-reviews-1,looping-reviews-2}.
Protein mediated DNA looping, for instance, is involved in gene regulatory processes
\cite{leduc14,metzler09pnas,cher11a}, or DNA and RNA constructs may be
used as molecular beacon sensors \cite{beacon}.

\begin{figure}
\begin{center}
\includegraphics[width=10cm]{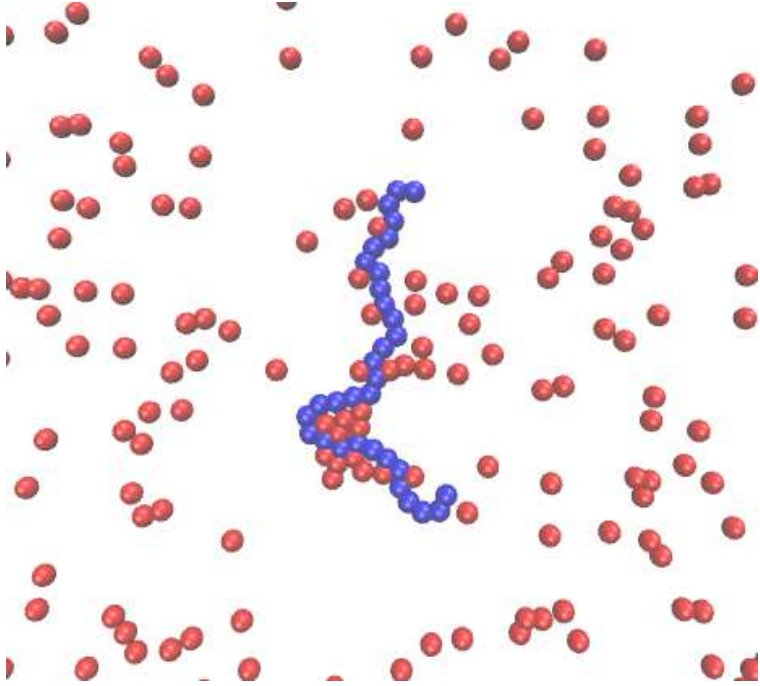}
\end{center}
\caption{Typical conformation of a polymer (blue chain) in a bath of active
particles (red discs) in two dimensions. Parameters: the polymerisation degree
of the chain is $n$=32, the packing fraction of SPPs is $\phi$=0.1, and the
particle activity is $F_a$=40 (see text for details). Video files for different
chain lengths and SPP activities are available in the Supplementary Material.}
\label{fig-schematic}
\end{figure}

We quantify the behaviour of a polymer chain immersed in a bath of SPPs in two
dimensions, see the snapshot of our systems in figure \ref{fig-schematic}. We
find that the activity $F_a$ of SPPs differently affects the chain conformations
depending on the chain bending stiffness $\kappa$. Concurrently, SPPs significantly
enhance the looping kinetics as well as give rise to a giant diffusivity of the
centre of mass motion of the chain due to SPP pooling in typical parachute like
chain configurations. We
analyse the diffusion of a monomeric tracer particle, which shows superdiffusive
motion on short time scales and Brownian behaviour with enhanced diffusivity in the
long time limit. In references \cite{kaiser14,cacciuto14} a similar system was
considered, the main focus being on equilibrium polymer properties such as the
gyration radius of the chain. Below we systematically analyse dynamic properties
of the polymer chain. Our results demonstrate that the equilibrium and dynamic
properties of polymers in an SPP bath are to be considered on the same footing.

This paper is organised as follows. We introduce our model and the simulations
methods in section \ref{sec-model}. In section \ref{sec-polymer-dimension} we
examine the equilibrium properties of the polymer chain. Section \ref{sec-looping}
presents the main results regarding the polymer looping properties. In section
\ref{sec-cm} we study the dynamical effects of SPPs on the tracer diffusion, in
order to understand its implications on the enhancement of the polymer looping
kinetics. We summarise our results and discuss their possible applications in
section \ref{sec-summary}.

\section{Model and Methods}
\label{sec-model}

To study the dynamics of a polymer chain immersed in a bath of SPPs, we employ
coarse grained computer simulations. The polymer chain is modelled as a bead
spring chain consisting of $n$ monomers of diameter $\sigma$, connected by
harmonic springs with the potential
\begin{eqnarray}
U_{\mathrm{s}}=\frac{k}{2}\sum_{i=2}^{n}\Big(|\mathbf{r}_{i}-\mathbf{r}_{i-1}|
-l_0\Big)^2.
\end{eqnarray}
Here $k$ is the spring constant and $l_0$ is the equilibrium bond length. The
self avoidance of the chain monomers is modelled by the repulsive part of the
Lennard-Jones (LJ) potential (the so called Weeks-Chandler-Andersen or WCA
potential),
\begin{eqnarray}
U_ {\mathrm{LJ}}(r)=4\epsilon\left[\left(\frac{\sigma}{r}\right)^{12}-\left(\frac{
\sigma}{r}\right)^6\right]+C(r_{\mathrm{cut}}).
\label{eq-lj}
\end{eqnarray}
for $r\leq r_{\mathrm{cut}}$, where $r_{\mathrm{cut}}$ is a cutoff length. Moreover
$C(r_{\mathrm{cut}})$ is a constant that ensures that $U_{\mathrm{LJ}}(r)=0$ for
separations $r>r_{\mathrm{cut}}$, and $r$ is the inter-monomer distance. The
potential strength is denoted by $\epsilon$. With the standard choice $r_{\mathrm{
cut}}=2^{1/6}\sigma$ for the cutoff length the potential is purely repulsive. In
what follows, we measure the length in units of $\sigma$ and the energy in units
of the thermal energy $k_{\mathrm{B}}T,$ where $k_{\mathrm{B}}$ is the Boltzmann
constant and $T$ is the absolute temperature. Below we set the model parameters to
$\sigma=1$, $l_0=1.12$, $k = 10^3$, and $\epsilon=1$. 

The bending energy of the chain is given by
\begin{eqnarray}
U_b=\frac{\kappa}{2}\sum_{i=2}^{n-1}\Big(\mathbf{r}_{i-1}-2\mathbf{r}_i+\mathbf{r}_{
i+1}\Big)^2,
\end{eqnarray}
where $\kappa$ is the bending stiffness. For a given value of $\kappa$, the chain
persistence length is $l_p\sim2\kappa l_0^3/(k_{\mathrm{B}}T)$ in two dimensions.
The end monomers are subject to short ranged attractive interactions with energy
$\epsilon_s$, mimicking the biologically relevant situation that closed structures
are energetically profitable, as known for specific DNA looping \cite{leduc14}
or closed single stranded DNA (hairpins) \cite{weiss13crowd}. We include the
attractive interactions via the LJ potential in Eq. (\ref{eq-lj}) but
with a larger cutoff distance and attraction strength $\epsilon_{s}$,
namely $$U_{\mathrm{att}}(r)=U_{\mathrm{LJ}}(r,\epsilon_{s})$$ and
$r_{\mathrm{cut}}=2\sigma$. The effects of the end-to-end stickiness on
the looping properties were considered by us recently \cite{shin15loop}.
In what follows we set $\epsilon_s =5k_{\mathrm{B}}T$. 

The dynamics of the position $\mathbf{r}_i(t)$ of the $i$th chain monomer is
described by the Langevin equation
\begin{eqnarray}
m\frac{d^{2}{\mathbf{r}}_i(t)}{dt^2}=-\boldsymbol\nabla [U_{\mathrm{s}}
+U_{\mathrm{LJ}}(r)+U_b]-\gamma\frac{d\mathbf{r}_i}{dt}+\boldsymbol{\xi}_i(t).
\label{eq-lang-passive}
\end{eqnarray}
Here $m$ is the monomer mass, $\gamma$ is its friction coefficient coupled to
the diffusivity via
\begin{equation}
D=k_{\mathrm{B}}T/\gamma,
\end{equation}
and $\boldsymbol{\xi}_{i}(t)$ represents a Gaussian white noise source of zero
mean with autocorrelation $\langle\boldsymbol{\xi}_i(t)\cdot\boldsymbol{\xi}_{i'}
(t')\rangle=4\gamma k_\mathrm{B}T\delta_{i,i'}\delta(t-t') $, where $\delta_{i,i'}$
is the Kronecker delta symbol.

The SPPs are modelled as disks of diameter $\sigma$ moving under the action of a
constant force along a predefined orientation vector
\begin{equation}
\mathbf{n}_j=\left\{\cos(\theta_j),\sin(\theta_j)\right\}.
\end{equation}
SPPs interact with each other as well as with polymer monomers via the WCA potential
(\ref{eq-lj}), and the position of each SPP is governed by the Langevin equation
\cite{cacciuto14}
\begin{eqnarray}
m\frac{d^{2}{\mathbf{r}}_{j}(t)}{dt^2}=-\boldsymbol\nabla U_{\mathrm{LJ}}(r)+F_a
\mathbf{n}_{j}(t)-\gamma\frac{d\mathbf{r}_j}{dt}+ \boldsymbol{\xi}_j(t).
\label{eq-lang-active}
\end{eqnarray}
Here $F_a$ is the active force amplitude, which is directly related to the SPP
propulsion strength: it can be expressed in terms of the P\'{e}clet number
$\mathrm{Pe}$ and the particle velocity $v$ in terms of
\begin{equation}
\mathrm{Pe}=\frac{v\sigma}{D}=\frac{F_a\sigma}{k_{\mathrm{B}}T}.
\end{equation}
The orientation $\theta_j$ of the velocity of the $j$th SPP is changing as function
of time according to the standard stochastic equation
\begin{eqnarray}
\dot{\theta} =\sqrt{2D_{r}}\times\xi_{r} (t). 
\end{eqnarray}
Here $\xi_{r}$ is the Gaussian white noise associated with the rotational
diffusivity $D_{r}$ which satisfies the relation $D_{r}=3D/\sigma^2$ (see, for
instance, reference \cite{cacciuto14}). Passive particles correspond to $F_{a}=0$,
the situation studied in the context of polymer looping with macromolecular
crowding in reference \cite{shin15loop}. 

\begin{figure*}
\begin{center}
\includegraphics[width=13.2cm]{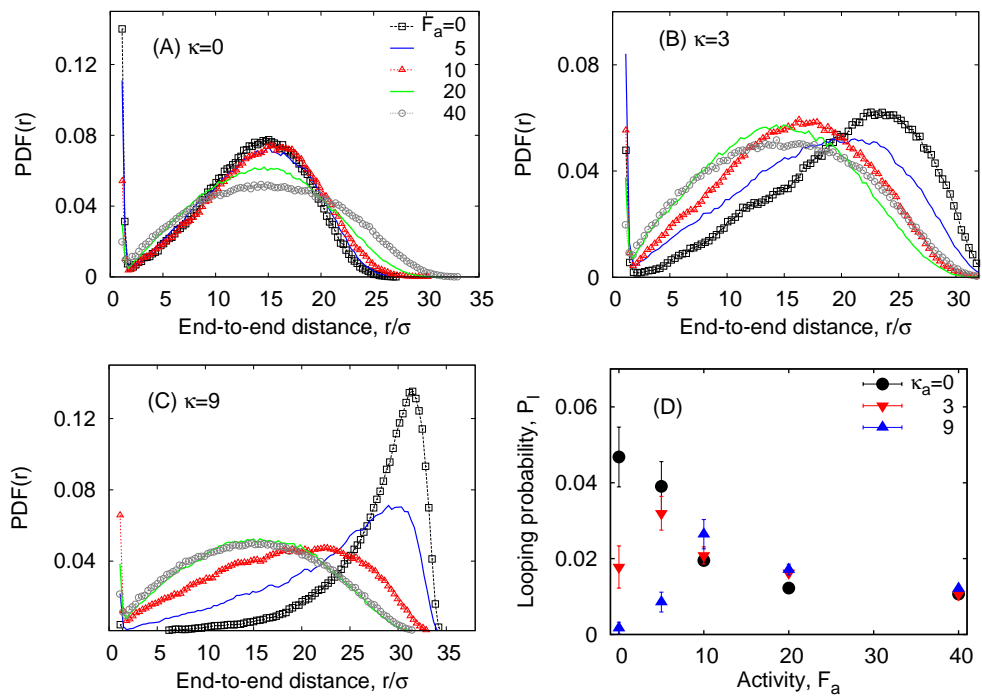}
\end{center}
\caption{Probability distribution of the polymer end-to-end distance for different
SPP activities $F_a$ and varying chain bending stiffness: (A) $\kappa=0$, (B)
$\kappa=3$, and (C) $\kappa=9$. In (D) the looping probability $P_l$ is plotted as
function of $F_a$ for $n=32$ and $\phi=0.05$.}
\label{fig-ete-distribution}
\end{figure*}

In our simulations we use periodic boundary conditions for a square box of area
$L^2$ where, depending on the length of the simulated chain, $L$ varies from 60
to 80. The packing fraction of SPPs is defined as $\phi=N_{\mathrm{cr}}A_{\mathrm{
cr}}/L^2$, where $N_{\mathrm{cr}}$ is the number of SPPs and $A_{\mathrm{cr}}=\pi(
\sigma/2)^2$ is the area of a single SPP. We consider rather dilute SPP systems
with $\phi\le0.1$. For both chain monomers and active particles we choose the unit
mass $m=1$ and a relatively large friction of $\gamma=5$ to ensure a quick momentum
relaxation. The time scale in the system is set by the elementary time $t_0=\sigma
\sqrt{m/(k_{\mathrm{B}}T)}$ \cite{allen}. We implement the Verlet velocity algorithm
\cite{allen} to simulate equation (\ref{eq-lang-passive}) and equation
(\ref{eq-lang-active}). The integration time step is $\Delta t=0.002$, and we
typically run $\sim10^{8\ldots9}$ steps to compute the quantities of interest.  

Generally the activity of SPPs may vary, or some particles in the bath may be
completely inactive. To account for this fact, in a part of our study below we
consider mixtures of active and inactive particles with respective fractions
$\phi_a$ and $\phi_i$. All these particles have the same mass and radius in the
simulations.

\section{Main Results}

\subsection{Polymer Dimensions}
\label{sec-polymer-dimension}

We first consider the probability distribution function (PDF) of the end to end
distance $r$ of the chain as extracted from long time computer simulations, see
figure \ref{fig-ete-distribution}A-C. In our simulations, due to the attraction
between the end monomers the standard PDF of the polymer \cite{cher11a} acquires
an additional peak around the minimum of the attraction potential at the end to
end distance $r\sim2^{1/6}\sigma$. For a flexible chain with $\kappa=0$ (figure
\ref{fig-ete-distribution}A) the chain gets more extended and the peak of the PDF
shifts to larger distances when the activity $F_a$ of SPPs increases. Conversely,
for semiflexible chains with a finite value of the polymer stiffness $\kappa>0$,
the peak is shifted to shorter distances (figure \ref{fig-ete-distribution}B,C).
These trends are similar to those of reference \cite{cacciuto14depletion}. 

This is the main effect of active particles on the static properties of passive
polymer chains in solutions. Inspecting snapshots of the simulations (see also
figure \ref{fig-schematic}) or the video files in the Supplementary Material,
one recognises that  active particles effect U or parachute like shapes of
the polymer. Such parachute shapes are also observed for membranous red blood
cells in cylindrical capillary flows and in blood vessels, see, e.g., reference
\cite{gompper-pnas}. For larger $F_a$ values, when the SPP forces are much larger
than the energetic scale for polymer bending, the PDF of the polymer end to end
distance is nearly independent on the chain stiffness $\kappa$.

\begin{figure}
\begin{center}
\includegraphics[width=12cm]{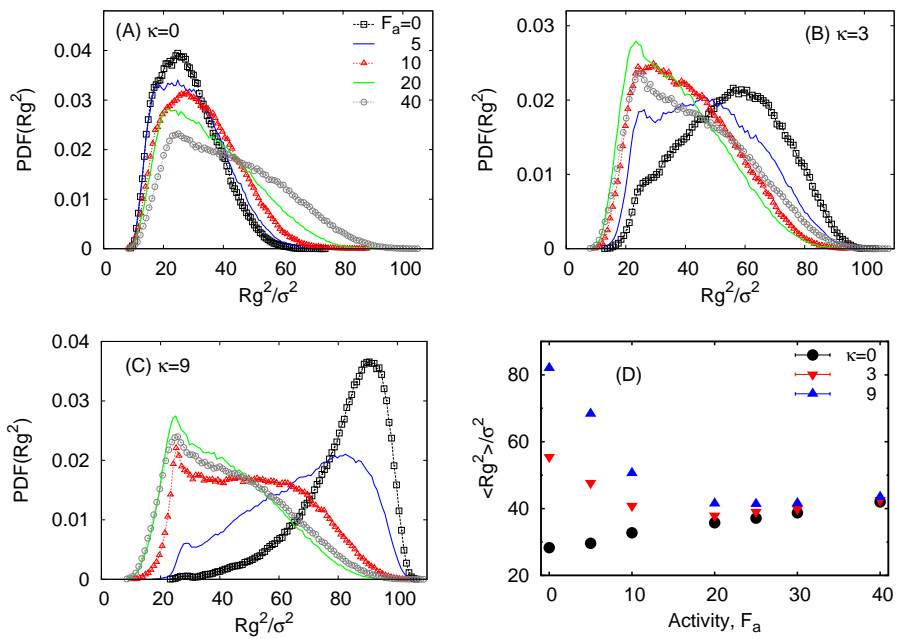}
\end{center}
\caption{Distribution of the gyration radius for varying $F_a$ (in each panel)
and bending stiffness (A) $\kappa=0$, (B) $\kappa=3$, and (C) $\kappa=9$. (D)
Average gyration radius as function of the SPP activity.
Parameters: $n=32$ and $\phi=0.05.$}
\label{fig-rg}
\end{figure}

In figure \ref{fig-rg} we also show the distribution of the radius of gyration
$R_g^{2}$ of the chain and its average value $\left< R_g^2\right>$. For flexible
chains the PDF of the gyration radius broadens towards larger values, causing
the monotonic increase of $\left< R_g^2\right>$. In contrast, for semiflexible
chains the gyration radius decreases for $F_a\leq20$, above this value it only
slightly increases, compare figure \ref{fig-rg}D. This behaviour is consistent
with the results of references \cite{kaiser14,cacciuto14}.

\subsection{Looping Probability and Looping Time}
\label{sec-looping}

The polymer end to end distance shows a highly erratic behaviour as function of
time (see also the movies in the Supplementary Material for chains of different
flexibility). The polymer ends tend to remain at short distances $\sim r_c$ due
to their attractive energy, while longer end to end distances with $r \sim r_{
\mathrm{eq}}$ are favourable entropically at equilibrium. We compute the looping
probability $P_l$ and the looping time $T_l$ from the time series of the polymer
end to end distance $r(t)$ generated in simulations. Similar to our recent studies
\cite{shin15loop,shin15loop2} the looping probability $P_l$ is defined as the
fraction of time during which the end to end distance of the chain is shorter than
$r_{c}$. In that sense the critical distance $r_{c}=1.75\sigma$ separates the
looped and un-looped states of the polymer, compare figure
\ref{fig-ete-distribution}A-C. 

Figure \ref{fig-ete-distribution}D shows the looping probability $P_l$ as function
of the activity $F_a$ of the SPPs. For flexible chains the value of $P_l$ decreases
monotonically with $F_a$. Conversely, for semiflexible polymers the looping
probability is non-monotonic in $F_a$. This observation indicates two competing
effects of the active particles: on the one hand SPPs increase the effective chain
flexibility resulting in higher $P_l$ values. On the other hand, SPPs facilitate
the unbinding of end monomers. We observe that, consistent with the shape of the
end to end distance PDF at large activity $F_a$ of SPPs in figures
\ref{fig-ete-distribution}A-C, for large $F_a$ the looping probability is almost
independent of $\kappa$, as demonstrated by figure \ref{fig-ete-distribution}D.

The polymer looping time $T_l$ is defined as the time interval within which the
distance $r$ reaches $r_{\mathrm{eq}}$ for the first time and the time it gets
shorter than the final distance $r_f=1.2\sigma$, details are shown in figure 3 of
reference \cite{shin15loop}. While the distances $r_c$ and $r_f$ are mainly
determined by the properties of the attractive potential of the end monomers, the
value of $r_{\mathrm{eq}}$ strongly varies with the chain length and the SPP
activity $F_a$. From the PDFs of the end to end distances we first determine
$r_{\mathrm{eq}}$ for a given chain length $n\sigma$ and particle activity and
then use them to compute the looping time $T_l$.

\begin{figure}
\begin{equation}
\includegraphics[width=8.8cm]{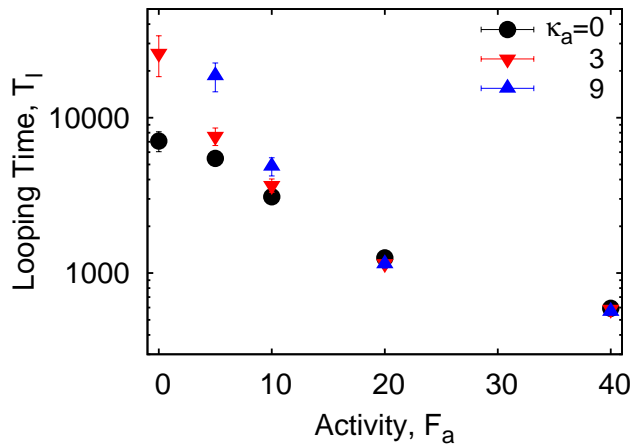}
\end{equation}
\caption{Polymer looping time versus SPP activity $F_a$ for polymer chains with
$n$=32 monomers. The simulation time of one point in this figure on a standard
workstation is around 6 h. The error bars for $P_l$
(figure \ref{fig-ete-distribution}D) and $T_l$ were computed
as the standard deviation of the mean via subdividing the time traces into
ten subsets.}
\label{fig-looping-time-activity}
\end{figure}

Although the effects of SPPs onto the spatial extension of the immersed polymers
were considered previously \cite{kaiser14,cacciuto14}, their dynamic
effects---particularly on the polymer end looping reaction---were not addressed in
detail. In figure \ref{fig-looping-time-activity} we present the polymer looping
times as function of the particle activity $F_a$. In free space or for $F_{a}=0$
the looping takes much longer for stiffer chains because of the large bending
energy required for a loop formation. As the SPP activity $F_{a}$ increases the
polymer looping time decreases, especially for stiff chains: we observe a reduction
of $T_l$ of more than two orders of magnitude, as evidenced in figure
\ref{fig-looping-time-activity}. Figure \ref{fig-ete-distribution} shows that for
large $F_a$ values the looping probability of flexible and semiflexible chains
behaves quite similarly, and figure \ref{fig-looping-time-activity} demonstrates
similar trends for the looping times $T_l$ of the polymer chains for large SPP
activities.

\begin{figure}
\begin{center}
\includegraphics[width=8cm]{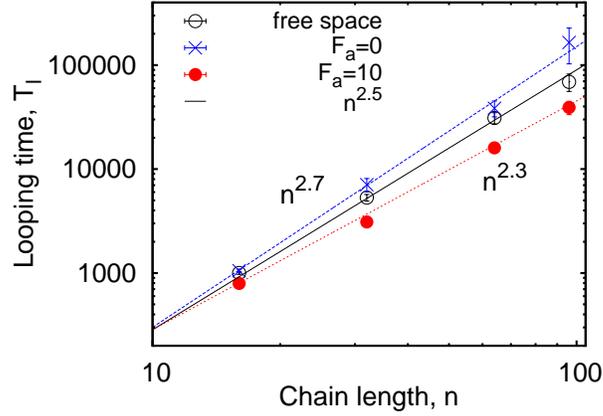}
\end{center}
\caption{Looping time versus chain length, plotted for the situation in free space
as well as for conditions of crowded inactive (blue) and active (red symbols)
particles. Parameters: $\phi$=0.05 and $\kappa=0$. The asymptotic behaviour of
equation (\ref{eq-loop-free-space}) is shown as the solid line, the slopes $2.2$
and $2.7$ are also shown.}
\label{fig-looping-time-length}
\end{figure}

Up to now, we only considered chains of length $n=32$ monomers. In figure
\ref{fig-looping-time-length} we now show the looping times as a function
of the chain length $n$. In free space ($\phi=0$), the looping time follows
the scaling behaviour \cite{shin15loop}
\begin{equation}
T_{l}(n)\sim n^{2\nu+1},
\label{eq-loop-free-space}
\end{equation}
with the Flory exponent $\nu=3/(d+2)=3/4$ for a polymer in two dimensions. With
$F_{a}=0$ (inactive crowders) the looping time increases for a given chain length
mainly due to a decreasing monomer diffusivity in the medium \cite{shin15loop}. In
the presence of active crowders the polymer kinetics becomes facilitated, especially
for longer chains, as shown by the red dots in figure \ref{fig-looping-time-length}.
Interestingly, in the presence of active particles, the scaling exponent of $T_l(n)$
decreases somewhat as compared to free space and passive crowders.

\begin{figure}
\begin{equation}
\includegraphics[width=8.8cm]{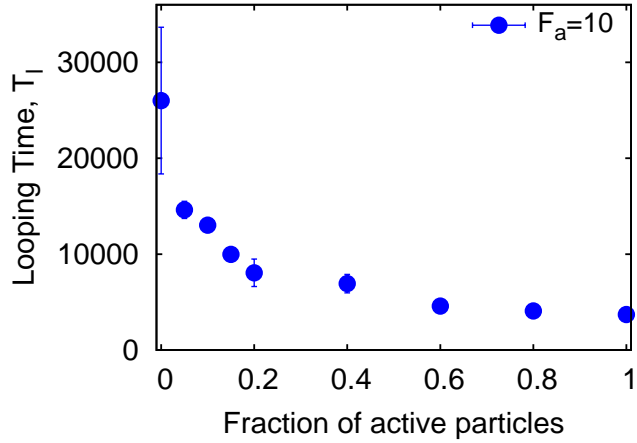}
\end{equation}
\caption{Polymer looping time versus fraction $\phi_a/\phi$ of SPPs at total crowder
fraction $\phi$=0.05. The chain length was $n=32$ and the bending stiffness
$\kappa=3$.}
\label{fig-looping-time-pp}
\end{figure}

In figure \ref{fig-looping-time-pp} we show the polymer looping time versus the
relative fraction of active particles, $\phi_a/\phi$ for the total crowding
fraction $\phi=\phi_a+\phi_i=0.05$. We observe that for small values $\phi_a/\phi$
the magnitude of $T_l$ initially drops sharply, while the decrease of $T_l$ for
larger
fractions of active particles is rather moderate. This indicates that the
transition from the non-active to the active results in figure
\ref{fig-looping-time-length} is rather non-uniform when active particles
are added into the solution.

\subsection{Tracer Diffusion, Polymer Diffusion and Monomer Displacements} 
\label{sec-cm}

To get a feeling for the effects of SPPs on the diffusion of passive particles
we now quantify the enhancement of the diffusive motion of a non-active tracer
in a bath of SPPs. We track a particle with diameter $\sigma$ (same size as the
monomers and SPPs) for varying SPP activities $F_a$. From the time series of the
tracer particle position $\mathbf{r}(t)=\{x(t),y(t)\}$ generated in our
simulations we calculate the time averaged mean squared displacement (MSD)
\cite{metz14,fran13}
\begin{eqnarray}
\overline{\delta^2_x(\Delta)}=\frac{1}{T-\Delta}\int_0^{T-\Delta}\Big[x({t'+
\Delta})-x({t'})\Big]^2dt',
\end{eqnarray}
where $\Delta$ is the so called lag time defining the width of the averaging
window shifted along the trajectory. Hereafter, the time averaged MSD is computed
with respect to one dimension only. The additional mean
\begin{equation}
\left<\overline{\delta^2(\Delta)}\right>=\frac{1}{N}\sum_{i=1}^N\overline{
\delta^2_i(\Delta)}
\end{equation}
over an ensemble of $N$ individual traces $\overline{\delta^2_i(\Delta)}$ will
be analysed below.

\begin{figure}
\begin{center}
\includegraphics[width=8cm]{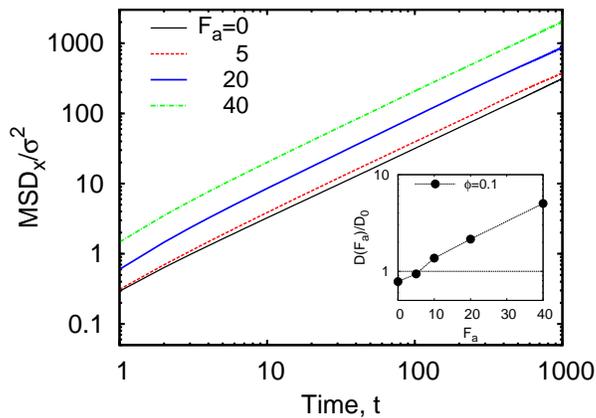}
\end{center}
\caption{Time averaged MSD $\left<\overline{\delta_x^2}\right>$ of a single tracer
at varying
SPP activities $F_a$ plotted for $\phi=0.05$. Inset: effective tracer diffusivity,
plotted in log-linear scale and normalised to the free space value.}
\label{fig-tracer-msd}
\end{figure}

We present the time averaged MSD of the tracer in figure \ref{fig-tracer-msd} for
varying SPP activity $F_a$. The time averaged MSD grows faster than for Brownian
motion (superdiffusion \cite{metz14}) only at very short times, $\Delta^{\star}
\lesssim2\ldots5$, and then turns into the linear Brownian scaling, as expected.
Since the momentum relaxation time, defined as $\sim m/\gamma=0.2$, is shorter
than the time scale $\Delta^{\star}$, the extended superdiffusion regime is
likely due to the impact of active particles.

\begin{figure}
\begin{center}
\includegraphics[width=7cm]{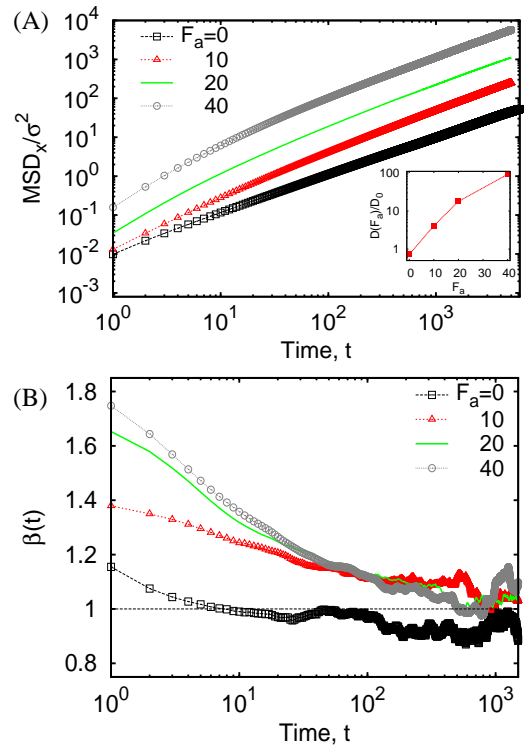}
\end{center}
\caption{(A) Time averaged MSD of the polymer centre of mass motion, averaged over
$N=5$ trajectories, as a function of
SPP activity. Inset: Effective diffusivity of the centre of mass motion, normalised
to the free-space value. (B) Local scaling exponent $\beta(t)$ of the
time averaged MSD. Parameters: $n$=32, $\kappa=0$, and $\phi=0.05.$} 
\label{fig-cm-motion}
\end{figure}

\begin{figure}
\begin{center}
A~\includegraphics[width=13cm]{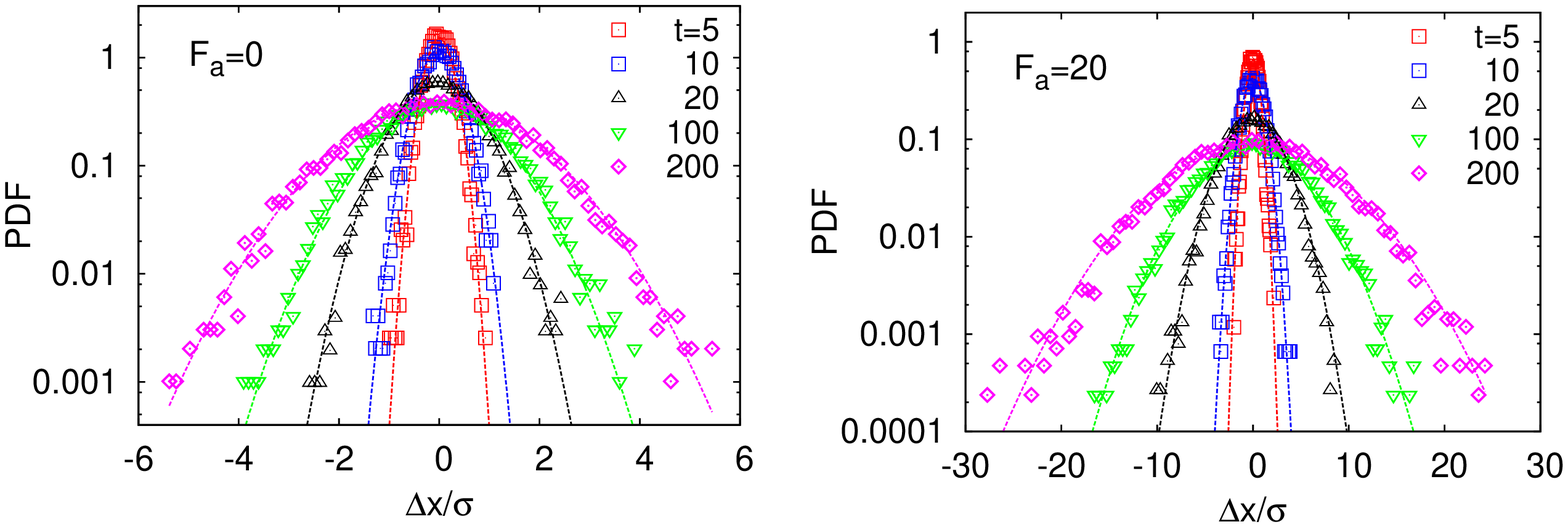}
B~\includegraphics[width=13cm]{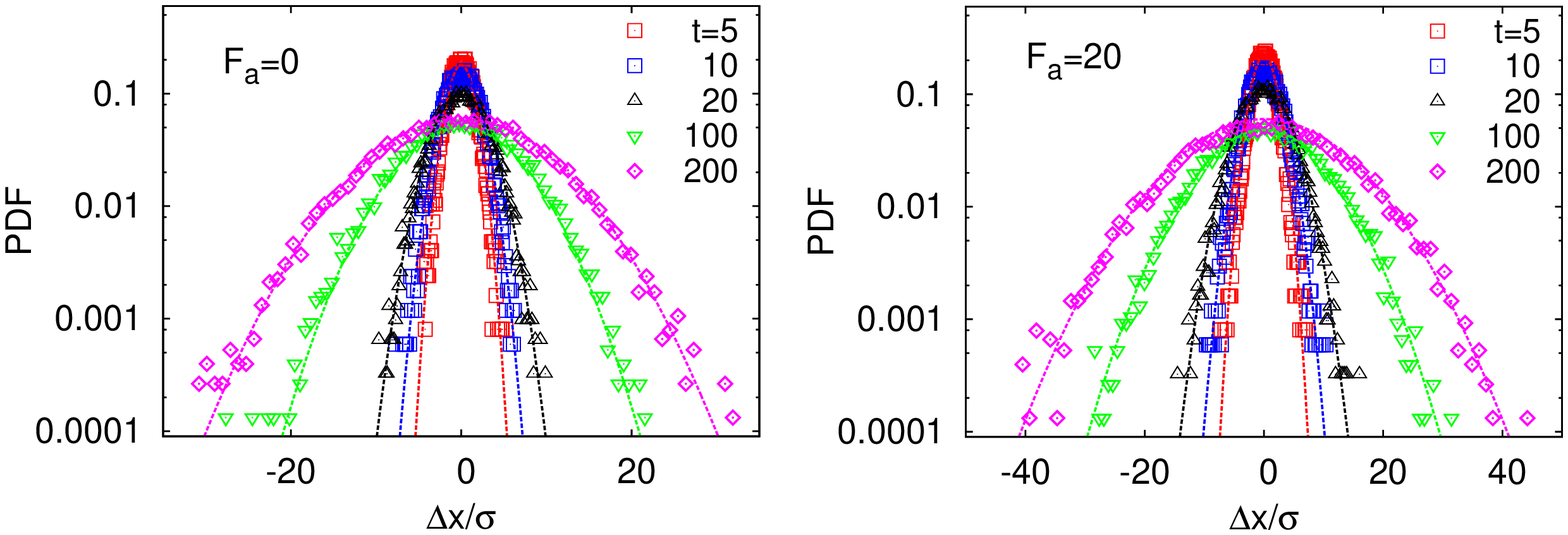}
\end{center}
\caption{PDFs of the centre of mass displacement of the polymer chain (A) and
tracer particle (B), plotted for a set of diffusion times $t$ and for two SPP
activities at crowding fraction $\phi$=0.05. The dotted lines are Gaussian fits.}
\label{fig-pdfs}
\end{figure}

We extract the diffusivity of the tracer particle through a linear fit to the
long time behaviour of the time averaged MSD in the range $\Delta=10^{2\ldots3}$.
As shown in the inset of figure \ref{fig-tracer-msd} in log-linear scale the
diffusivity increases almost exponentially with $F_a$. This enhancement is the
main reason of the dramatic facilitation of the polymer looping kinetics by
highly active particles, as demonstrated in figure
\ref{fig-looping-time-activity} as function of the SPP activity $F_a$. This is
one of the main conclusions of the current paper.

Similarly, in figure \ref{fig-cm-motion}A we show the time averaged MSD of the
polymer chain's centre of mass for different SPP activities $F_a$. Comparison of
the magnitudes of the time averaged MSDs shows that, as expected, the centre of
mass diffusion of the entire chain is evidently much slower than that of a single
tracer particle. In figure \ref{fig-cm-motion}B we compute the local scaling
exponent of the MSD \cite{metz14,fran13}
\begin{equation}
\nonumber
\beta(\Delta)=\frac{d\log\left<\overline{\delta^2(\Delta)}\right>}{d\log(\Delta)}.
\end{equation}
We observe that at short time scales the MSD increases superdiffusively with $\beta
>1$ and the anomalous diffusion exponent grows with increasing $F_a$ values. At
longer times the diffusion exponent decreases and around $\Delta\simeq10^3$ the
motion becomes nearly Brownian, albeit with an enhanced diffusivity at higher $F_a$
values. 

In the inset of figure \ref{fig-cm-motion}A we show the chain diffusivity of the
centre of mass motion as
function of $F_a$, normalised with respect to the value in free space. The
enhancement of the diffusivity of the polymer centre of mass is nearly two orders
of magnitude, that is much higher than that of a monomeric tracer particle shown
in the inset of figure \ref{fig-tracer-msd}. This stronger enhancement is due
to pooling of SPPs in the concave region of the parachute-shaped polymer chain,
resulting in directed motion and faster diffusion of the polymer. This remarkable
finding of a giant diffusivity enhancement is our second major result.

We also show the PDFs of the displacement of the polymer chain and of the tracer
particle in panels A and B of figure \ref{fig-pdfs}, respectively. Both for active
and inactive crowders the PDFs exhibit Gaussian profiles, see the dotted fits in
the figure. In the presence of active particles, the width of the corresponding
displacement PDF becomes wider, consistent with the enhanced diffusivity. This is
particularly clear for the polymer centre of mass displacement shown in panel A of
figure \ref{fig-pdfs}. Interestingly, even at short times---when the time averaged
MSD shows a superdiffusive scaling---the PDFs remain nearly Gaussian.

\section{Conclusions}
\label{sec-summary}

Actively driven systems are inherently out of equilibrium and exhibit peculiar
behaviours, for instance, in the ratcheting of motors \cite{ratchetmotor10}, the
formation of living crystals \cite{livingcrystal13,speck13prl}, phenomena of
ordering \cite{stark}, mesoscale
turbulence phenomena \cite{turbulence12}, or superfluidic behaviour may be
observed in bacterial suspensions \cite{super}. Even elementary laws of
thermodynamics may no longer hold \cite{cacciuto14pressure,solon14,solon15}.
In that sense the behaviour of active liquid systems is as rich as that of
active soft matter \cite{review15}.

We studied the dynamics of a polymer chain in a bath of SPPs using Langevin
dynamics simulations in two dimensions. We first considered the equilibrium
behaviour of the gyration radius, the end to end distance distribution, and
the looping probability of the chain as function of the activity of the SPPs.
We found that a flexible polymer extends monotonically with the activity. In
contrast, for a semiflexible chain---due to a competition of the the chain bending
and active forces---the polymer size varies non-monotonically with the particle
activity. For a larger activity of SPPs---when active forces become dominant
over the chain elasticity effects---the extension of both semiflexible and
flexible chains behaves quite similarly. 

SPPs also significantly impact the polymer kinetics, the focus of this study.
Overall, due to the enhanced diffusivity of the chain monomers, the looping
dynamics becomes
faster. Especially for the case of stiffer chains the presence of SPPs renders
the chains effectively softer, and the looping kinetics is dramatically enhanced
in comparison to that of flexible chains. Our results indicate that the activity
of a cell medium, as mimicked above by active particles, may indeed facilitate
DNA loop formation, effectively making the molecule more flexible. 

Examining the motion of a tracer particle in comparison to the motion of the centre
of mass of the polymer chain in the bath of SPPs, we observe a giant diffusivity
for the driven polymer. We ascribe this to the parachute like shape of the polymer
in the SPP bath. Due to the accumulation of SPPs in the concave region of the chain
the
polymer performs an extended ballistic motion over time scales, that are considerably
longer than that of a single monomer. The chain motion at long times becomes
Brownian, but with an unexpectedly high diffusivity. Interestingly, even at
time ranges on which the time averaged MSD is superdiffusive, the distribution
of the particle displacement remains Gaussian. This result differs from experimental
observations of an extended exponential decay of the displacement of a tracer in
swimming microorganisms \cite{lipchaber00}. It would thus be interesting to check
whether incorporation of hydrodynamic interactions would reproduce such non-Gaussian
behaviour in the model of SPP baths. Moreover, it is an interesting questions
whether the effect of SPP pooling and the ensuing giant diffusivity enhancement
of the polymer motion also arises in three dimensions.

Our analysis here was performed with an \emph{in vitro\/} bath of SPPs in mind. In
particular the crowding fraction was chosen to be fairly low. Inside living cells
active particles such as molecular motors drive the environment out of equilibrium
and fluctuating forces inside cells may indeed become an order of magnitude larger
than at the conditions of thermal equilibrium \cite{gallet09}. Concurrently, the
metabolic cell activity significantly affects the nature of the cytoplasm
\cite{cytoplasm14}. However, the (macromolecular) crowding fraction in cells
typically is of the order of $30\ldots35\%$ \cite{zimmerman93,elcock10} and thus
exceeds the value of our simulations by far. Moreover, these crowders are quite
complex macromolecular objects, which tune the reaction kinetics stability of
biopolymers \cite{pincusPRL15}. It will therefore be interesting
to extend our study to higher crowder fractions and different particle geometries,
such as star shapes \cite{mercedes}.

\ack

We acknowledge funding from the Academy of Finland (Suomen Akatemia, Finland
Distinguished Professorship to RM) and the Deutsche Forschungsgemeinschaft
(DFG Grant to AGC).


\section*{References}

\begin{thebibliography}{99}

\bibitem{review12b} P. Romanczuk, M. B\"ar, W. Ebeling, B. Lindner, and L.
Schimansky-Geier, Eur. Phys. J. Spec. Topics \textbf{202}, 1 (2012), and
references therein.

\bibitem{review13} M. C. Marchetti, J. F. Joanny, S. Ramaswamy, T. B. Liverpool,
J. Prost, M. Rao, and R. A. Simha, Rev. Mod. Phys. \textbf{85}, 1143 (2013).

\bibitem{julicher-motors} F. J\"ulicher, A. Ajdari, and J. Prost, Rev. Mod. Phys.
\textbf{69}, 1269 (1997); A. B. Kolomeisky and M. E. Fisher, Ann. Rev. Phys. Chem.
\textbf{58} 675 (2007); R. D. Astumian and P. H{\"a}nggi, Phys. Today
\textbf{55}(11), 33 (2002).

\bibitem{igor} D. Robert, T. H. Nguyen, F. Gallet, and C. Wilhelm, PLoS ONE
\textbf{4}, e10046 (2010); I. Goychuk, V. O. Kharchenko, and R. Metzler, PLoS ONE
\textbf{9}, e91700 (2014); Phys. Chem. Chem. Phys. \textbf{16}, 16524 (2014).

\bibitem{elgeti-microswimmers} J. Elgeti, R. G. Winkler, and G. Gompper, Rep. Prog.
Phys. \textbf{78}, 056601 (2015), and references therein.

\bibitem{review12} M. E. Cates, Rep. Prog. Phys. \textbf{75}, 042601 (2012).

\bibitem{peshehody} I. Karamouzas, B. Skinner, and S. J. Guy, Phys. Rev. Lett.
\textbf{113}, 238701 (2014); D. Brockmann, L. Hufnagel, and T. Geisel, Nature,
\textbf{439}, 462 (2006); O. B{\'e}nichou, C. Loverdo, M. Moreau and R. Voituriez,
Rev. Mod. Phys., \textbf{83}, 81 (2011); P. C. Bresloff, J. M. Newby, Rev.
Mod. Phys. \textbf{85}, 135 (2013); C. Song, T. Koren, P. Wang,
and A.L. Barab\'{a}si, Nature Physics \textbf{6}, 818 (2010).

\bibitem{golestanian07} J. R. Howse, R. A. L. Jones, A. J. Ryan, T. Gough, R.
Vafabakhsh, and R. Golestanian, Phys. Rev. Lett. \textbf{99}, 048102 (2007);
W. F. Paxton et al., J. Am. Chem. Soc. \textbf{128}, 14881 (2006);
J. Palacci, C. Cottin-Bizonne, C. Ybert, and L. Bocquet,
Phys. Rev. Lett. \textbf{105}, 088304 (2010);

\bibitem{sana10} H.-R. Jiang, N. Yoshinaga, and M. Sano, Phys. Rev. Lett.
\textbf{105}, 268302 (2010).

\bibitem{peter} P. H{\"a}nggi and F. Marchesoni, Rev. Mod. Phys. \textbf{81},
387 (2009); P. K. Ghosh, P. H{\"a}nggi, F. Marchesoni, and F. Nori, Phys. Rev.
E \textbf{89}, 062115 (2014).

\bibitem{aljaz} C. Loverdo, O. B{\'e}nichou, M. Moreau, and R. Voituriez,
Nat. Phys. \textbf{9}, 134 (2008); J. Stat. Mech. P02045 (2009); A. Godec and
R. Metzler, Phys. Rev. E \textbf{92}, 010701(R) (2015); K. Chen, B. Wang, and
S. Granick, Nature Mat. \textbf{14}, 589 (2015).

\bibitem{search} G. M. Viswanathan, M. G. E. da Luz, E. P. Raposo, and H. E.
Stanley, The Physics of Foraging. An Introduction to Random Searches and
Biological Encounters (Cambridge University Press, New York, 2011);
M. A. Lomholt, T. Koren, R. Metzler, and J. Klafter, Proc. Natl. Acad. Sci. USA
\textbf{105}, 11055 (2008) V. V. Palyulin, A. V. Chechkin, and R. Metzler,
Proc. Natl. Acad. Sci. USA \textbf{111} 2931 (2014); D. W. Sims et al., Nature
\textbf{451}, 1098 (2008).

\bibitem{lipchaber00} X.-L. Wu and A. Libchaber, Phys. Rev. Lett. \textbf{84},
3017 (2000).

\bibitem{poon13} A. Jepson, V. A. Martinez, J. Schwarz-Linek, A. Morozov, and
W. C. K. Poon, Phys. Rev. E \textbf{88}, 041002(R) (2013).

\bibitem{goldstein09} K. C. Leptos, J. S. Guasto, J. P. Gollub, A. I. Pesci, and
R. E. Goldstein, Phys. Rev. Lett. \textbf{103}, 198103 (2009).

\bibitem{marenduzzo14} A. Morozov and D. Marenduzzo, Soft Matter \textbf{10},
2748 (2014).

\bibitem{haim} M. Mussel, K. Zeevy, H. Diamant, and U. Nevo, Biophys. J.
\textbf{106}, 2710 (2014); S. Roy et al, J. Neurosci. \textbf{27}, 3131
(2007); D. A. Scott et al., Neuron. \textbf{70}, 441 (2011).

\bibitem{ao58} S. Asakura and F. Oosawa, J. Chem. Phys. \textbf{22}, 1255 (1954).

\bibitem{cacciuto14depletion} J.  Harder, S. A. Mallory, C. Tung, C. Valeriani,
and A. Cacciuto, J. Chem. Phys. \textbf{141}, 194901 (2014).

\bibitem{ni15} R. Ni, M. A. Cohen Stuart, and P. G. Bolhuis, Phys. Rev. Lett.
\textbf{114}, 018302 (2015).

\bibitem{kaiser14} A. Kaiser and H. L\"owen, J. Chem. Phys. \textbf{141}, 044903
(2014).

\bibitem{cacciuto14} J. Harder, C. Valeriani, and A. Cacciuto, Phys. Rev. E
\textbf{90}, 062312 (2014).

\bibitem{looping-reviews-1} R. Schleif, Annu. Rev. Biochem. \textbf{61}, 199 (1992).

\bibitem{looping-reviews-2} K. S. Matthews, Microbiol. Mol. Biol. Rev. \textbf{56},
123 (1992).

\bibitem{leduc14} C. Tan, S. Saurabh, M. P. Bruchez, R. Schwartz, and P. LeDuc,
Nature Nanotech. \textbf{8}, 602 (2013); J. M. G. Vilar and I. Saiz, ACS Synth.
Biol. \textbf{2}, 576 (2013).

\bibitem{metzler09pnas} B. van den Broek, M. A. Lomholt, S.-M. J. Kalisch, R.
Metzler, and G. J. L. Wuite, Proc. Natl. Acad. Sci. U.S.A. \textbf{105}, 15738
(2008); M. A. Lomholt, B. v. d. Broek, S.-M. J. Kalisch, G. J. L. Wuite,
and R. Metzler, Proc. Natl. Acad. Sci. USA \textbf{106}, 8204 (2009);
M. Bauer, E. S. Rasmussen, M. A. Lomholt, and R. Metzler, Sci. Rep. \textbf{5},
10072 (2015); D. Swigon, B. D. Coleman, and W. K. Olson, Proc. Natl.
Acad. Sci. U.S.A. \textbf{103}, 9879 (2006);
A. Amitai and D. Holcman, Phys. Rev. Lett. \textbf{110}, 248105 (2013).

\bibitem{cher11a} A. G. Cherstvy, Europ. Biophys. J. \textbf{40}, 69 (2011);
A. G. Cherstvy, J. Phys. Chem. B \textbf{115}, 4286 (2011).

\bibitem{beacon} S. Tyagi and F. R. Kramer, Nat. Biotechnol. \textbf{14}, 303 (1996);
G. Bonnet, O. Krichesvky, and A. Libchaber, Proc. Natl. Acad. Sci. USA \textbf{95},
8602 (1998).

\bibitem{weiss13crowd} O. Stiehl, K. Weidner-Hertrampf, and M. Weiss, New J. Phys.
\textbf{15}, 113010 (2013).

\bibitem{shin15loop} J. Shin, A. G. Cherstvy, and R. Metzler, Soft Matter
\textbf{11}, 472 (2015).

\bibitem{allen} M. P. Allen and D. J. Tildesley, \textit{Computer Simulation of
Liquids} (Clarendon, Oxford, 1994).

\bibitem{gompper-pnas} H. Noguchi and G. Gompper, Proc. Natl. Acad. Sci. U. S. A.
\textbf{102}, 14159 (2005).

\bibitem{shin15loop2} J. Shin, A. G. Cherstvy, and R. Metzler, ACS Macro Lett.
\textbf{4}, 202 (2015).

\bibitem{metz14} R. Metzler, J.-H. Jeon, A. G. Cherstvy, and E. Barkai, Phys. Chem.
Chem. Phys. \textbf{16}, 24128 (2014). 

\bibitem{fran13} F. H\"ofling and T. Franosch, Rep. Prog. Phys. \textbf{76},
046602 (2013).

\bibitem{ratchetmotor10} R. Di Leonardo et al., Proc. Natl. Acad. Sci. U. S. A.
\textbf{107}, 9541 (2010).

\bibitem{livingcrystal13} J. Palacci, S. Sacanna, A. F. Steinberg, D. J. Pine,
and P. M. Chaikin, Science \textbf{339}, 936 (2013).

\bibitem{speck13prl} I. Buttinoni, J. Bialke, F. Kuemmel, H. L\"owen, C. Bechinger,
and T. Speck, Phys. Rev. Lett. \textbf{110}, 238301 (2013).

\bibitem{stark} M. Hennes, K. Wolff, and H. Stark, Phys. Rev. Lett. \textbf{112},
238104 (2014).

\bibitem{turbulence12} H. H. Wensink, J. Dunkel, S. Heidenreichd, K. Drescher,
R. E. Goldstein, H. L\"owen, and J. M. Yeomans, Proc. Natl. Acad. Sci. U. S. A.
\textbf{109}, 14308 (2012).

\bibitem{super} H. M. L{\'o}pez, J. Gachelin, C. Douarche, H. Auradou, and E.
Cl{\'e}ment, Phys. Rev. Lett. \textbf{115}, 028301 (2015).

\bibitem{cacciuto14pressure} S. A. Mallory, A. Saric, C. Valeriani, and A. Cacciuto,
Phys. Rev. E \textbf{89}, 052303 (2014).

\bibitem{solon14} A. P. Solon, Y. Fily, A. Baskaran, M. E. Cates, Y. Kafri, M.
Kardar, and J. Tailleur, arXiv: 1412.3952.

\bibitem{solon15} A. P. Solon, J. Stenhammar, R. Wittkowski, M. Kardar, Y. Kafri,
M. E. Cates, and J. Tailleur, Phys. Rev. Lett. \textbf{114}, 198301 (2015).

\bibitem{review15} A. M. Menzel, Phys. Rep. \textbf{554}, 1 (2015).

\bibitem{gallet09} F. Gallet, D. Arcizet, P. Bohec, and  A. Richert,   
Soft Matter \textbf{5}, 2947 (2009).

\bibitem{cytoplasm14} B. R. Parry, I. V. Surovtsev, M. T. Cabeen, C. S. O'Hern,
E. R. Dufresne, and C. Jacobs-Wagner, Cell \textbf{156}, 183 (2014).

\bibitem{zimmerman93} S. B. Zimmerman and A. P. Minton, Annu. Rev. Biophys. Biomol.
Struct. \textbf{22}, 27 (1993).

\bibitem{elcock10} A. R. McGuffee and A. H. Elcock, PLoS Comp. Biol. \textbf{6},
e1000694 (2010).

\bibitem{pincusPRL15} H. Kang, P. A. Pincus, C. Hyeon, and D. Thirumalai, Phys.
Rev. Lett. \textbf{114}, 068303 (2015); H. X. Zhou, G. Rivas, and A. P. Minton,
Annual Rev. Biophys. \textbf{37}, 375 (2008); N. A. Denesyuk and D. Thirumalai,
J. Am. Chem. Soc. \textbf{133}, 11858 (2011); N. M. Toan, D. Marenduzzo, P. R.
Cook, and C. Micheletti, Phys. Rev. Lett. \textbf{97}, 178302 (2006); A. P. Minton,
Curr. Opin. Struct. Biol. \textbf{10}, 34 (2000); M. Yanagisawa, T. Sakaue, and K.
Yoshikawa, Intl. Rev. Cell \& Molec. Biol. \textbf{307}, 175 (2014); A. R. Denton,
Intl. Rev. Cell \& Molec. Biol. \textbf{307}, 27 (2014).

\bibitem{mercedes} J. Shin, A. G. Cherstvy, and R. Metzler, E-print
arXiv:1507.01176.

\end{thebibliography}
\end{document}